\documentclass[twocolumn,showpacs,showkeys, preprintnumbers,amsmath,amssymb]{revtex4-1} 

 \usepackage[utf8]{inputenc}
 \usepackage[T1]{fontenc}

 \PassOptionsToPackage{hyphens}{url} 
 \usepackage[colorlinks=true, pdfborder={0 0 0}]{hyperref}
 \usepackage{url}

 \makeatletter\newcommand\multiabel[1]{\quad \refstepcounter{equation}(\theequation)\ltx@label{#1}}\makeatother

\usepackage{graphicx}
\usepackage{xcolor}
\usepackage{enumitem}
\usepackage{empheq}
\usepackage{bbold}
\usepackage{dsfont}

\pdfpageattr{/Group << /S /Transparency /I true /CS /DeviceRGB>>} 

\def\thesection{\arabic{section}}
\def\thesubsection{\arabic{section}.\arabic{subsection}}

\makeatletter
\renewcommand{\p@subsection}{}
\renewcommand{\p@subsubsection}{}
\makeatother

\renewcommand{\vec}{\mathbf}

\usepackage{ulem}
\definecolor{darkgreen}{rgb}{0,0.5,0} 
\definecolor{violet}{rgb}{0.5,0,0.5}
\definecolor{orange}{rgb}{0.8,0.5,0.2}
\definecolor{gray}{rgb}{0.3,0.3,0.3}
\definecolor{green}{RGB}{50,177,65}

\begin{document}


\title{Topological phase transition in coupled rock-paper-scissor cycles}

\author{Johannes Knebel$^{1, \dagger}$, Philipp M. Geiger$^{1, \dagger}$, and Erwin Frey$^{1}$}

\email[]{frey@lmu.de}
\thanks{$\dagger$ J.K. and P.M.G. contributed equally to this work.}
\date{\today}

\affiliation{Arnold-Sommerfeld-Center for Theoretical Physics and Center for NanoScience,
Department of Physics, Ludwig-Maximilians-Universität München, D–80333 Munich, Germany}

\begin{abstract}
A hallmark of topological phases is the occurrence of topologically protected modes at the system’s boundary.
Here we find topological phases in the antisymmetric Lotka-Volterra equation (ALVE). The ALVE is a nonlinear dynamical system and describes, e.g., the evolutionary dynamics of a rock-paper-scissors cycle.
On a one-dimensional chain of rock-paper-scissor cycles, topological phases become manifest as robust polarization states. At the transition point between left and right polarization, solitonic waves are observed. 
This topological phase transition lies in symmetry class $D$ within the ``ten-fold way'' classification as also realized by 1D topological superconductors.

\end{abstract}

\maketitle

\twocolumngrid

\textbf{Introduction.} 
Topological phases were discovered in condensed matter physics~\cite{Klitzing1980,Thouless1982, Haldane1988,Hasan2010, Chiu2016} and recently extended to classical physics such as topological mechanical metamaterials~\cite{Kane2014,Paulose2015, VitelliIrvine2015,Susstrunk2016, Chen2016, Fan2019,  Kedia2020}.
From a phenomenological point of view, topological phases are paramount for the following characteristics~\cite{Hasan2010, Chiu2016}: (i)~\textit{Localization}--dynamical excitations become localized at the system's boundary; (ii)~\textit{Robustness}--these boundary modes are robust against perturbations of the system's parameters and noise; (iii)~\textit{Phase transition}--at the transition point between the topological phases, the dynamical mode expands throughout the whole system.
From a theoretical point of view, topological phases are determined by nontrivial topological properties of the system's bulk, which are classified in terms of the system's symmetries. These properties give rise to gapless boundary modes and explain the observations~(i)-(iii)~\cite{Hasan2010,Chiu2016,Susstrunk2016}.

Thus far, the study of topological phases and their realization in soft-matter and biological systems has only started to develop~\cite{Prodan2009, Souslov2017, Delplace2017, Shankar2017, Murugan2017, Dasbiswas2018, Zhou2018, Souslov2019, Sone2019, Pedro2019, Scheibner2020, Yamauchi2020, Yoshida2020}. As to what extent topological phases may determine the behavior of dynamical systems that arise, for example, in population dynamics was, however, not addressed. 
Ultimately, it would be interesting to design biological set-ups with nontrivial topological properties so that one obtains robust dynamical modes with the above characteristics (i)-(iii).

In this work, we make a step in this direction by showing that topological phases can be realized with the antisymmetric Lotka-Volterra equation (ALVE). The ALVE is a paradigmatic model for studying coexistence and survival in population dynamics~\cite{Volterra1931, Goel1971} and also describes the condensation of non-interacting bosons in driven-dissipative set-ups~\cite{Vorberg2013, Knebel2015}. In population dynamics, the ALVE governs, for example, the evolutionary dynamics of the rock-paper-scissors game, in which each of the three strategies dominates one strategy and is dominated by another one, such that all strategies survive~\cite{Hofbauer1998, Reichenbach2006, Claussen2008, Szolnoki2014}. 

\begin{figure}[htp!]
\centering
\includegraphics{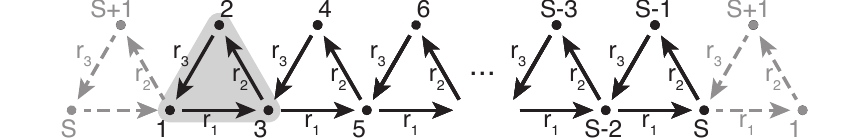}
\caption{
\textbf{One-dimensional chain of rock-paper-scissors cycles.} The interactions on the $S$ sites of the RPS chain (one single RPS cycle highlighted) are captured by the antisymmetric matrix~$A$ in Eq.~\eqref{eq:Matrix_An}. An arrow from one site to another indicates that mass is transported in this direction at rate $r_1, r_2,r_3>0$ following the ALVE~\eqref{eq:ALVE}; the skewness $r= r_2/r_3$ defines the control parameter.
The auxiliary site $S+1$ facilitates periodic boundary conditions (dashed lines) within the framework of topological band theory.
}
\label{fig:model}
\end{figure}
\begin{figure*}[th!]
\centering
\includegraphics[width=1.0\textwidth]{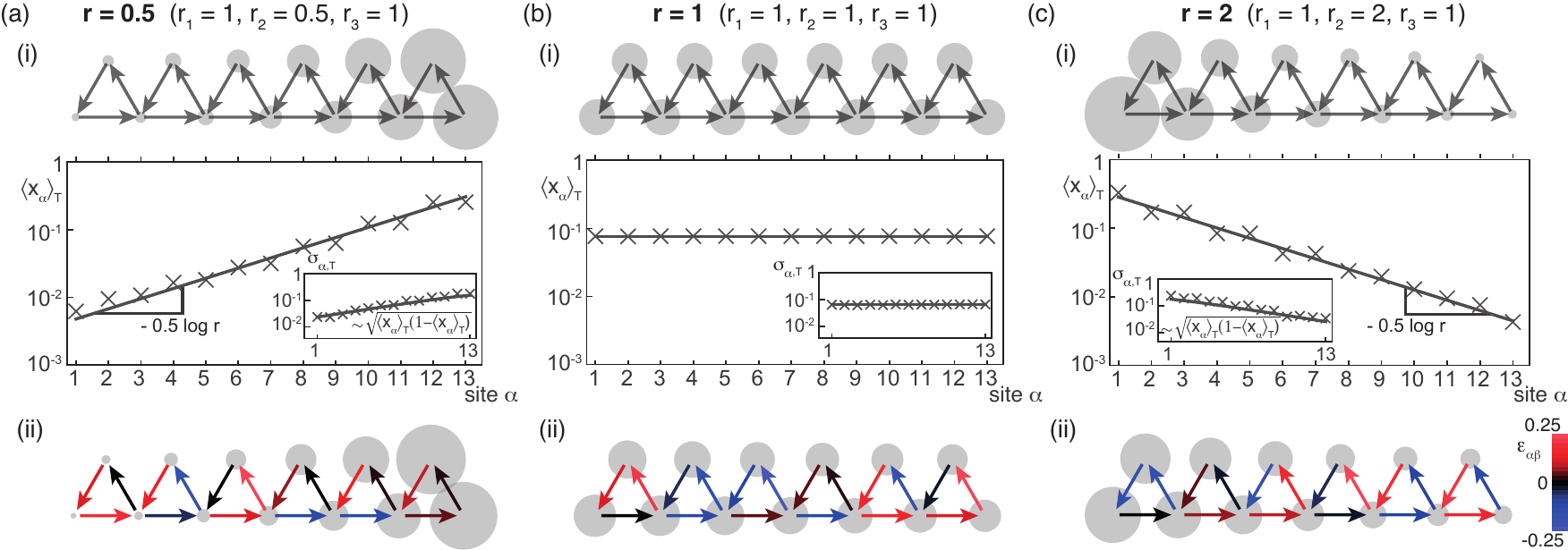}
\caption{\textbf{Polarization of mass to the boundary depending on the parameter $r$.}
Time averages of the masses in the stationary state from single realizations of the ALVE~\eqref{eq:ALVE} are depicted on the RPS chain (disk size encodes magnitude) and on lin-log scale for $T=5000$; see (a-c)(i). Mass becomes polarized to the boundary (to the right for $r<1$, (a)(i), and to the left for $r>1$, (c)(i)) in that $\langle x_\alpha\rangle_T$ decays exponentially into the bulk at a length scale set by $\ln{r}$ for all initial conditions. For $r=1$, mass becomes equally distributed on the whole chain (b)(i). Polarization is no state of rest as shown by the non-vanishing fluctuations $\sigma_{\alpha, T}$ around the average masses (insets of graphs (a-c)(i)).
Polarization is robust against perturbations~(a-c)(ii). Positive matrix entries are perturbed as $a_{\alpha\beta}'=a_{\alpha\beta}(1+\epsilon_{\alpha\beta})$, $\epsilon_{\alpha\beta}$ is uniformly sampled in the interval $[-0.25, 0.25]$ (encoded by color of the links). Polarization to the boundary emerges with the same characteristics as for the unperturbed set-up.
}
\label{fig:polarization}
\end{figure*}

Mathematically speaking, the ALVE is a nonlinear, mass-conserving dynamical system defined on $S$ sites. The mass at each site~$\alpha$ is denoted as $x_\alpha$, which evolves according to the coupled ordinary differential equations:
\begin{align}
\label{eq:ALVE}
\partial_t x_\alpha = x_\alpha \textstyle\sum_{\beta = 1}^{S} a_{\alpha\beta} x_\beta\ ,\quad \alpha = 1\dots, S\ .
\end{align}
The real-valued $S\times S$ matrix $A = \{ a_{\alpha\beta}\}$ is antisymmetric ($a_{\alpha\beta} = -a_{\beta\alpha}$) and defines how mass is transported between two sites in a nonlinear interaction $\sim x_\alpha x_\beta$. 

Here we study the ALVE on a one-dimensional chain of coupled rock-paper-scissors cycles (``RPS chain'', see Fig.~\ref{fig:model}). We observe behaviors that resemble key features of a topological phase transition: (i) mass becomes polarized to the right or left boundary of the RPS chain independent of the initial conditions; (ii) polarization is robust against perturbations of the model parameters; (iii)~at the transition point between left and right polarization, the overall mass expands throughout the whole RPS chain and, moreover, solitary waves are observed. 
To explain these dynamics, we relate the polarization states of the ALVE~\eqref{eq:ALVE} to properties of the antisymmetric matrix $A$. We show that the RPS chain encompasses a gap in the spectrum of $A$ and an intrinsic ``particle-hole symmetry'' and, thus, falls into the symmetry class $D$ within the ``ten-fold way'' classification scheme of gapped free-fermion systems~\cite{Chiu2016}. Hence, left and right polarization  on the RPS chain are distinguished by a $\mathbb{Z}_2$ invariant, which characterizes the observed topological phase transition. 
Intriguingly, the topological polarization states arise as an entirely nonlinear phenomenon that cannot be rationalized within the framework of linear wave theory.
We envision that topological phases in the ALVE could be constructed in two dimensions as well and that the described mechanism might guide one path to design robust topological phases in nonlinear dynamical systems, which could be realized in biological set-ups.

\textbf{Model.} 
The RPS chain is composed of RPS cycles coupled in one dimension (see Fig.~\ref{fig:model}) and  represented by the antisymmetric matrix of size $S=2n-1$ ($n>1$),
\begin{align}\label{eq:Matrix_An}
A = 
\begin{psmallmatrix}
0 & r_3 & - r_1  & 0 & 0 &\dots  & 0& 0& 0 \\
-r_3 & 0 & r_2 & 0 &  0 &\dots & 0& 0& 0\\
r_1& -r_2& 0 & r_3 & -r_1 &\dots  & 0& 0& 0 \\
0& 0& -r_3 & 0 & r_2 &\dots & 0& 0& 0\\
0& 0& r_1 & -r_2 & 0 &\dots & 0& 0& 0\\
\vdots & \vdots &  \vdots &  \vdots & \vdots & \ddots & \vdots & \vdots & \vdots\\
0 & 0 & 0 & 0  & 0 &\dots & 0 & r_3 & -r_1\\
0 & 0 & 0 & 0  & 0 &\dots & -r_3  & 0& r_2 \\
0 & 0 & 0 & 0  & 0 &\dots & r_1  & -r_2  & 0\\
\end{psmallmatrix}\ ,
\end{align}
with rate constants $r_1,r_2,r_3>0$. In our numerical simulations of the ALVE~\eqref{eq:ALVE}, time is rescaled such that $r_1=1$. The ratio $r:= r_2/r_3$ serves as the control parameter for the dynamics and is referred to as \textit{skewness}. 
The RPS chain can be thought of as a one-dimensional chain of nonlinear oscillators because each isolated RPS cycle represents a local oscillator in which mass oscillates between the different sites. For $r\neq 1$, mass is skewed towards certain sites within a single oscillating RPS cycle. 

The initial masses in the ALVE~\eqref{eq:ALVE} are normalized ($\sum_{\alpha} x_{0,\alpha} = 1$) and strictly positive ($x_{0,\alpha} > 0$ for all~$\alpha$). 
Due to the antisymmetry of $A$, the total mass is conserved for all times $t \geq 0$ ($\sum_\alpha \partial_t x_\alpha = 0$). 
Furthermore, all masses remain bounded away from~0 ($x_\alpha\geq const>0$, for all $\alpha$) for all times for any choice of rates $r_1,r_2,r_3>0$~\cite{Knebel2015, Geiger2018}.
In the context of evolutionary game theory, this means that all strategies coexist on the RPS chain.

\textbf{Phenomenology.} 
In the numerical simulations of the RPS chain, we observed a surprisingly rich dynamics for how the mass distributes depending on the skewness~$r$. 

(i) \textit{Localization.} 
For skewness $r< 1$, the overall mass on the RPS chain becomes polarized to the right boundary over time irrespective of the initial conditions (Fig.~\ref{fig:polarization}(a)(i)), whereas for $r>1$, mass polarizes to the left (Fig.~\ref{fig:polarization}(c)(i)). 
Polarization on the RPS chain becomes manifest as an exponential decay of the average mass per site from the boundary into the bulk.
We quantify this polarization by averaging the mass at every site over a time window $T \gg 1$, $\langle x_\alpha \rangle_T = 1/T \int_0^T\text{d} t\; x_\alpha (t) $ (Fig.~\ref{fig:polarization}(a,c)(i)).
We observed that, for $r\neq 1$, average masses decay from the boundary into the bulk as $\langle x_\alpha \rangle_{T} \sim \exp (- \alpha/l_p)$ for $\alpha\geq 1$, numerically consistent with $l_p = 2/ \ln r$ as the penetration depth.
Such polarization arises for any initial mass distribution and is already visible for a small system size of $S=13$ (Fig.~\ref{fig:polarization}). 

The ALVE~\eqref{eq:ALVE} is a deterministic dynamical system approaching a stationary state at large times. This stationary state can be characterized by the average mass per site, $\langle x_\alpha \rangle_T$ (see polarization above) and the fluctuations around the averages within a framework of thermodynamic equilibrium~\cite{Goel1971}.
To quantify the fluctuations at site $\alpha$, we measured the standard deviation $\sigma_{\alpha,T}=\sqrt{\langle x_\alpha^2\rangle_T -\langle x_\alpha\rangle_T^2}$; see insets in Fig.~\ref{fig:polarization}(a,c)(i). 
Interestingly, the standard deviation scales with the values of the average masses themselves as $\sigma_{\alpha,T}\sim \sqrt{\langle x_\alpha \rangle_T(1-\langle x_\alpha \rangle_T)}$, while their precise amplitude is determined by the initial conditions~\cite{Supplement}. Taken together, also the fluctuations in the polarization state scale universally in that they decay exponentially into the bulk from the boundary at which the mass is localized.

(ii) \textit{Robustness.} 
Polarization is robust against perturbations of the model parameters. 
For example, Fig.~\ref{fig:polarization}(a,c)(ii) illustrates the polarization of mass to the boundary when the positive matrix entries~\eqref{eq:Matrix_An} are perturbed as $a_{\alpha\beta}'=a_{\alpha\beta}(1+\epsilon_{\alpha\beta})$. Here, all $\epsilon_{\alpha\beta}$ are independently sampled from the same probability distribution whose mean $\langle \epsilon \rangle$ fulfils $|\langle \epsilon \rangle| \leq 2/S$ (with $S \gg 1$)~\cite{Supplement}.
As another example for the robustness of polarization, we found that mass becomes localized to the boundary of the chain when the coupling between even sites is introduced, that is, the links between sites $2m$ and $2(m-1)$ for $m= 2, \dots, n-1$ are added to the RPS chain.
Even with these additional couplings does the dynamics proceed into the same qualitative polarization states as without perturbation of the ALVE~\eqref{eq:ALVE}~\cite{Supplement}.

\begin{figure}[t!]
\centering
\includegraphics{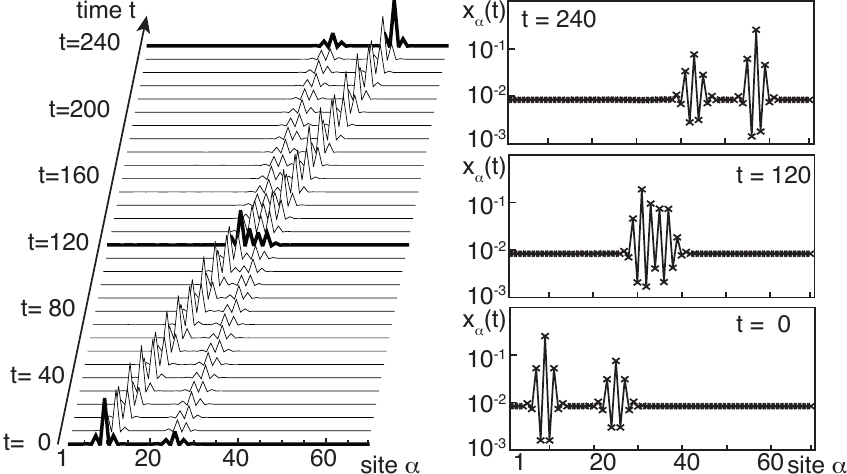}
\caption{\textbf{Solitary waves at the transition point $r=1$.}
The suitably prepared wave package ($t=0$) remains localized and travels along the RPS chain ($S=80$) without changing its shape. In an interaction with another solitary wave ($t=120$), the shapes of both wave packages remain unchanged afterwards ($t=240$). The initial wave packages were numerically obtained from the dispersion of a single mass peak.
}
\label{fig:soliton}
\end{figure}

(iii) \textit{Phase transition.} 
For $r=1$, the average masses expand throughout the whole chain, marking the transition point between the two polarization states. In the stationary state, the overall mass is on average uniformly distributed on the chain with $\langle x_\alpha \rangle_T =1/S$ for all $\alpha$. More generally, from our numerical simulations it turns out that it is not possible to tune any set of rate constants on the RPS chain such that one passes from polarization at one boundary to polarization at the other boundary without crossing a transition point at which the average masses expand throughout the whole chain.

Furthermore, we observed solitary mass waves at the transition point $r=1$. Mass packages that are suitably initialized at a few neighboring sites show soliton-like properties (see Fig.~\ref{fig:soliton}): They are spatially confined; their shape does not change; and after an interaction with other solitary waves, their shape and speed remains unchanged~\cite{Remoissenet1999, ablowitz1991}.
It will be interesting to further characterize these solitary waves and connect them to already known solitons in similar set-ups~\cite{Zakharov1974, Manakov1975, Suris2003, Yan2006, Chen2014, Chaunsali2019}.

\textbf{Analysis.} 
Taken together, the combination of the observations on (i) localization, (ii) robustness, and (iii) phase transition share characteristic features of a topological phase transition that underlies the  behavior of the ALVE~\eqref{eq:ALVE} on the RPS chain. In the following, we make this hypothesis rigorous. 
First, we outline how fixed points $\vec{x}^*$ ($\partial_t\vec{x}|_{\vec{x}^*} = 0$) of the ALVE are determined by strictly positive kernel vectors of $A$. 
Second, we show that the qualitative changes in the dynamics of the ALVE can be understood in terms of an underlying topological phase transition, which is derived from the bulk properties of $A$ within the framework of topological band theory.
Third, we compute the kernel vector for the RPS chain by means of a graph-theoretical interpretation of the Pfaffian of $A$ and thereby confirm the results obtained from the topological band theory approach.

\textit{Fixed points of the RPS chain.} 
For the RPS chain defined by the matrix~$A$ in Eq.~\eqref{eq:Matrix_An}, there exists a unique, strictly positive kernel vector $\vec{c}$ for every choice of parameters ($A\vec{c}=\vec{0}$ with $c_\alpha>0$ for all $\alpha$ and $\sum_\alpha c_\alpha =1$).
This vector $\vec{c}$ gives rise to the fixed point $\vec{x}^*=\vec{c}$ of the ALVE~\eqref{eq:ALVE}, and no further fixed points with $x^*_\alpha>0$ for all $\alpha$ exist~\cite{Supplement}.
The existence of a unique vector $\vec{c}$ derives from the cyclic structure of the single RPS cycles that are concatenated in one dimension and its explicit form is given below in Eq.~\eqref{eq:kernel_vector}.
If the ALVE~\eqref{eq:ALVE} is initialized away from this fixed point $\vec{x}_0\neq\vec{x}^*= \vec{c}$, the masses $\vec{x}$ do not approach the fixed point, but instead are confined on trajectories around the value of $\vec{c}$. It turns out that the average masses are given by $\langle x_\alpha\rangle_\infty = c_\alpha$ for all initial conditions~\cite{Knebel2015}. 
This way, the long-time dynamics of the ALVE on the RPS chain (average masses $\langle \vec{x}\rangle_\infty$) are determined by algebraic properties of $A$ (kernel vector~$\vec{c}$).

\textit{Topological band theory of the RPS chain.} 
To further characterize the algebraic properties of $A$, the RPS chain is extended by the auxiliary site $S+1$ and periodic boundary conditions (PBC) are employed (Fig.~\ref{fig:model}). The corresponding antisymmetric matrix, $A_\text{PBC}$, is of size $2n$ and block-circulant (that is, translationally invariant)~\cite{Davis1994, Gray2006}. To relate our approach to condensed matter physics, we define the ``RPS Hamiltonian'' $H := iA_\text{PBC}$ ($i$ denotes the imaginary unit), which is a Hermitian matrix ($H^\dagger = H$) with only real eigenvalues. Below, we analyze $H$ in the framework of topological band theory~\cite{Ashcroft1976, Kane2013}.

\begin{figure}[tb!]
\centering
\includegraphics{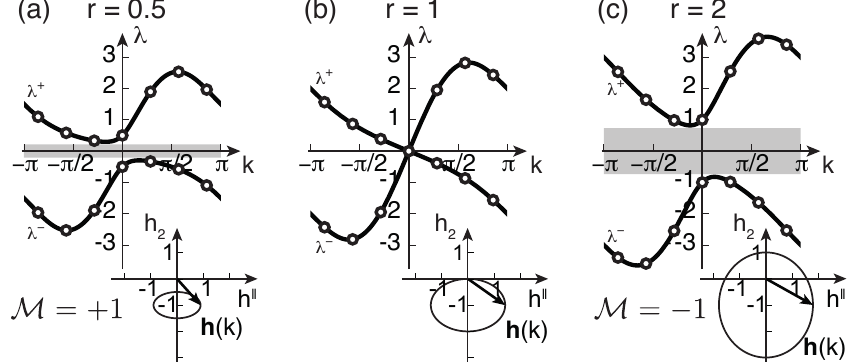}
\caption{\textbf{Topological phase transition in the RPS chain.} 
The band structure of the RPS Hamiltonian $H$ is point-symmetric for all values of $r$ (white dots denote the eigenvalues of $H$ for size $S=14$) and exhibits a spectral gap around 0 for $r\neq 1$ (gray shade).
The band structures for $r<1$ and $r>1$ are topologically distinct in how the eigenvectors change within the Brillouin zone, which is quantified by the different values of the topological invariant $\mathcal{M}$ and visualized by whether the projection of the vector $\vec{h}(k)$ winds around 0 \cite{Supplement}. Thus, a topological phase transition occurs at $r=1$.
}
\label{fig:spectrum}
\end{figure}

The spectrum of $H$ is characterized by its band structure. Starting from the eigenvalue equation for $H$ and exploiting translational invariance, the eigenvectors of $H$ can be decomposed into a plane wave part and a within-cell alignment part $\tilde{\vec{u}}(k)$, which fulfils the much simpler eigenvalue equation $\lambda(k)\tilde{\vec{u}}(k) = \tilde{H}(k)\tilde{\vec{u}}(k)$ for the Fourier-transformed Hamiltonian $\tilde{H}$; $k$ denotes the wave number in the Brillouin zone ($k = \frac{2\pi}{n} l$ and $l= \lfloor \frac{n}{2}\rfloor, \dots,  0, 1, \dots, \lfloor \frac{n}{2}\rfloor-1$).
For the RPS chain, one finds,
\begin{align}\label{eq:Hamiltonian}
\tilde{H}= 
\begin{psmallmatrix}
2  r_1 \sin k & - r_2 \sin k + i(r_3 - r_2 \cos k) \\- r_2 \sin k-i(r_3 - r_2 \cos k)   & 0
\end{psmallmatrix}\ ,
\end{align}
which can be written as $\tilde{H}(k) = \vec{h}(k)\cdot \vec{\sigma}$, with $\vec{h}(k) = (h_0(k), h_1(k), h_2(k), h_3(k))= (r_1 \sin k, -r_2 \sin k, -r_3+r_2 \cos k, r_1 \sin k)$;  $\vec{\sigma} = (\sigma_0, \sigma_1, \sigma_2, \sigma_3)$ denotes the Pauli matrices~\cite{Arfken1985} with $\sigma_0$ as the $2\times 2$ identity matrix~\cite{Supplement}.

How the spectral and topological properties of $H$ depend on the control parameter $r$ can be derived from Eq.~\eqref{eq:Hamiltonian}; see Fig.~\ref{fig:spectrum}. First, the spectrum of $H$ exhibits two bands of eigenvalues, $\lambda^{+}$ and $\lambda^{-}$, on the Brillouin zone $k\in [-\pi, \pi)$ since the unit cell is constituted of two sites ($2m+1 \to 2m$). For $r\neq 1$, the two bands are separated by a spectral gap that closes only for $r=1$ at $k=0$. 

Second, the spectrum of $H$ is point-symmetric with respect to the origin, $\lambda^{+}(k) = -\lambda^{-}(-k)$. This property follows from the intrinsic ``particle-hole symmetry'' of $H$ because it is defined by a real-valued antisymmetric matrix. In formal terms, $H$ fulfils the operator identity $\mathcal{C}\tilde{H}(k)\mathcal{C}^{-1} = -\tilde{H}(-k)$, with $\mathcal{C}:=\sigma_0\circ \kappa$ and $\kappa$~as~the complex conjugation operator. However, $H$ does not have time-reversal or chiral symmetry.
Thus, the RPS Hamiltonian $H$ falls into symmetry class $D$ in one dimension within the ``ten-fold way'' classification scheme of gapped free-fermion systems~\cite{Chiu2016}. In fact, $H$ can be interpreted as a Bogoliubov-de Gennes mean-field description of superconductivity in the Majorana representation~\cite{Kitaev2001, Ryu2002, Budich2013, Chiu2016}.

Gapped Hamiltonians in the symmetry class $D$ in 1D admit topological phases that are characterized by a $\mathbb{Z}_2$ invariant~\cite{Kitaev2001, Budich2013, Chiu2016}. This invariant, $\mathcal{M}$, is the sign of the Pfaffian of $A_\text{PBC}$ and can be computed from $\tilde{H}$ as $\mathcal{M} = \text{sign}(1-r^2)$~\cite{Budich2013}. Thus, a topological phase transition occurs at $r=1$: for $r<1$, the RPS Hamiltonian $H$ is in the ``topologically trivial phase'' ($\mathcal{M} = +1$); for $r>1$, the ``topologically non-trivial phase'' is attained ($\mathcal{M} = -1$); see insets of Fig.~\ref{fig:spectrum} for an illustration. In other words, the two phases ($r<1$ and $r>1$) are topologically distinct in that they cannot be smoothly deformed into one another without closing the spectral gap. 

Through the so-called bulk-boundary correspondence~\cite{Kane2014, Chiu2016, Tauber2020}, topological properties of the bulk (the periodic RPS chain with $S+1$ sites) become manifest at the boundary of the open system (the original RPS chain with $S$ sites).
More precisely, upon removing the auxiliary site $S+1$ and returning to the open RPS chain of size $S$, the spectral gap for $r\neq 1$ is populated by a zero eigenvalue with a corresponding, topologically protected, strictly positive kernel vector, whose entries are localized at the boundary of the RPS chain~\cite{Supplement}. This bulk-boundary correspondence is made rigorous by introducing a Toeplitz matrix as an intermediary between the two matrices $A$ and $A_\text{PBC}$ and applying the Szegö-Widom theorem~\cite{Basor2019,Dubail2017,Deift2012}.
We conclude that the polarization states of the ALVE~\eqref{eq:ALVE} on the RPS chain correspond to gapless boundary modes. Thus, left and right polarization constitute topologically distinct stationary states of the ALVE, which cannot be transformed into each other without passing through the phase transition at $r=1$.

\textit{Kernel vectors of the RPS chain.} 
Finally, we briefly present the exact form of the strictly positive kernel vector $\vec{c}$ of $A$. 
To determine $\vec{c}$, we employed the graph-theoretical interpretation of the Pfaffian, such that the kernel of $A$ is related to the network representation of $A$ in Fig.~\ref{fig:model}; see details in references~\cite{Geiger2018, Supplement}. 
As a result, the kernel vector can be written as ($\vec{c} = (c_1, \dots, c_{2n-1})$):
\begin{align}\label{eq:kernel_vector}
\begin{pmatrix}
c_{2m-1} \\ 
c_{2m}
\end{pmatrix}
= \frac{1}{C}
\begin{pmatrix}
r_2\\ 
r_1
\end{pmatrix}
r^{-(m-1)} \ ,\quad m=1, \dots, n-1\ ,
\end{align}
and $c_{2n-1} = r_3^{n-1}r_2^{n-2}/C$; $C$ denotes the normalization constant and ensures $\sum_\alpha c_\alpha = 1$. Thus, average masses $\langle \vec{x} \rangle_\infty = \vec{c}$ decay as $c_\alpha\sim \exp{(-\alpha/l_p)}$ with penetration depth $l_p=2/ \ln{r}$ from the boundary (analogously for $r<1$); see Fig.~\ref{fig:polarization}. 
This explicit construction of the kernel vector agrees with the result obtained within the approach of topological band theory and, thus, confirms the topological nature of the transition at $r=1$.

\textbf{Discussion.} 
In this work, we found a topological phase transition in the stationary state of the ALVE~\eqref{eq:ALVE} on the RPS chain. Stationary states are linked to strictly positive kernel vectors of the defining antisymmetric matrix. These kernel vectors are topologically protected and give rise to the gapless boundary modes that are observable as robust polarization of mass on the one-dimensional chain. Notably, these topological phases are entirely nonlinear in that they cannot be understood as the superposition of linear waves.

We envision that the results of this work could be extended to specific higher dimensions. In two dimensions, the symmetry class $D$ within the ``ten-fold way'' classification scheme admits topological phases characterized by the Chern number, whereas in three dimensions no topological phase transition occurs~\cite{Chiu2016}. In two dimensions, such a topological phase should become observable as a unidirectional flow of mass at the system's boundary. A possible lattice candidate might be constructed as a two-dimensional carpet of RPS cycles as the natural extension of the RPS chain.

Beyond the observation of topological phases in the ALVE, it might be possible to generalize the approach of this work to other dynamical systems in biological physics whose attractors are nonlinear oscillators or limit cycles~\cite{Novak2008}. Here we employed the ALVE on a RPS cycle as the constituting building block, but other other local oscillators may serve equally well. By suitably coupling these oscillators in the spirit of this work, we believe that topological phases as robust dynamical modes in biological systems could be designed.

\textbf{Acknowledgement.} 
We thank Chase Broedersz and Isabella Graf for discussions at early stages of this project, and Anton Souslov for discussions. 
J.K. thanks the Boulder Summer School 2015 through which this work was initially inspired.
This research was supported by the Excellence Cluster ORIGINS, which is funded by the Deutsche Forschungsgemeinschaft (DFG, German Research Foundation) under Germany's Excellence Strategy – EXC-2094 – 390783311, and the funding initiative ``What is life?'' of the VolkswagenStiftung.
The authors declare no conflict of interest. 

\normalem
\bibliography{PRL_references.bib}

\clearpage
\onecolumngrid


\renewcommand\thefigure{S\arabic{figure}}  
\renewcommand\thesection{S\arabic{section}}
\renewcommand\thesubsection{\thesection.\arabic{subsection}}
\renewcommand{\theequation}{\thesection.\arabic{equation}}
\setcounter{page}{1}
\setcounter{figure}{0}
\setcounter{equation}{0}

\section*{Supplementary information}
\section{Stationary states on the RPS chain}

\subsection{Computation of the strictly positive kernel vector of \texorpdfstring{$A$}{\textit{A}}} \label{sec:kernel}

As mentioned in the main text, we compute the strictly positive kernel vector of $A$ for the RPS chain of size $S=2n-1$ via the so-called adjugate vector of an antisymmetric matrix. To this end, the notion of the Pfaffian of an antisymmetric matrix and its graph-theoretical interpretation are exploited and outlined in this subsection. The details of this approach, in particular statements (1) and (2) below, are explained in reference~\cite{Geiger2018}.

\textit{(1) The kernel of an antisymmetric matrix is characterized via its adjugate vector.}
Because $S=2n-1$ is odd, at least one eigenvalue of $A$ is zero and, thus, the kernel of $A$ contains at least one independent kernel vector. To further characterize the kernel of $A$, one may employ the algebraic tool of the so-called adjugate vector $\vec{v}\in\mathbb{R}^{S}$ as follows: If $\vec{v}=\vec{0}$, then the kernel of $A$ has dimension bigger than 1; if $\vec{v}\neq\vec{0}$, then the kernel of $A$ is one-dimensional and $\vec{v}$ is the kernel vector upon normalization. In other words, the kernel vector of $A$ can be directly computed via the adjugate vector (if it is non-zero); see~\cite{Geiger2018} for details. 
The entries of the adjugate vector are computed as follows:
\begin{align}\label{eq:adjugate_vector_general}
    v_i = (-1)^{\alpha +1}\text{Pf}(A_{\hat{\alpha}})\ , \ \alpha =1,\dots , S\ .
\end{align}
Here, $A_{\hat{\alpha}}\in\mathbb{R}^{(S-1) \times (S-1)}$ denotes the even-sized antisymmetric matrix that is obtained by removing both the $\alpha$-th row and column from $A$, and $\text{Pf}(A_{\hat{\alpha}})$ denotes the Pfaffian of $A_{\hat{\alpha}}$, which is the determinant-like function for antisymmetric matrices with the property that the Pfaffian squared equals the determinant.\\

\textit{(2) Graph-theoretical interpretation of the Pfaffian.} 
The Pfaffian of odd-sized antisymmetric matrices is zero (because the kernel is non-trivial). For even-sized matrices, the Pfaffian is defined through a combinatorial formula in close relation to the determinant. Alternatively, the Pfaffian may be defined in a graph-theoretical manner based on the network that corresponds to the antisymmetric matrix (see Fig.~\ref{fig:model}). We briefly outline the graph-theoretical definition for the purpose of our analysis in the following; see~\cite{Geiger2018} for details and examples.

The Pfaffian of an antisymmetric matrix $A$ (of even size $2n$, $n=1, 2, \dots$) may be calculated as the sum of all signed  perfect matchings of the network corresponding to $A$. A perfect matching $\mu$ of a network is a subset of the network's edges such that all nodes are covered exactly once.  
With this notion, the Pfaffian is computed as:
\begin{align}\label{eq:Pfaffian_matchings}
\text{Pf} (A) = \textstyle\sum_{\mu}\left(\text{sign}(\sigma_{\mu }) \prod_{(\alpha\to \beta)\in \mu } a_{\alpha\beta} \right) \ ,
\end{align}
in which the sum runs over all perfect matchings $\mu$ of the network of $A$ (an edge
$\alpha \to \beta$ contributing to a perfect matching $\mu$ contributes with $a_{\alpha\beta} < 0$ to the product), and the permutation $\sigma_{\mu}=(\alpha_1\ \beta_1\ \alpha_2\  \beta_2\ \dots  \alpha_n\ \beta_n)$ denotes the partition of the nodes that is obtained from the edges in the perfect matching $\mu$. The sign of the perfect matching $\mu$, $\mathrm{sign}(\sigma_{\mu })$, is determined by the number of transpositions needed to permute the partition $(\alpha_1, \beta_1, \alpha_2,  \beta_2, \dots,  \alpha_n, \beta_n)$ into the ordered partition of nodes $(1, 2, \dots, 2n)$.\\

\begin{figure*}[th!]
\centering
\includegraphics[width=1.\textwidth]{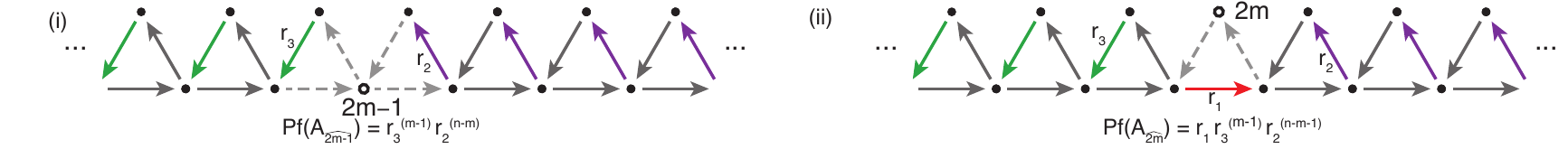}
\caption{\textbf{Calculation of the adjugate vector of $A$ for the RPS chain.}
(i) The $(2m-1)$-th component of the adjugate vector, $v_{2m-1}$, is calculated as the Pfaffian of the subnetwork obtained by deleting node $(2m-1)$ and determining all perfect matchings of the resulting network. For the RPS chain, only one perfect matching contributes as indicated by the green and purple arrows. (ii) Analogously, for the $2m$-th component, only one perfect matching contributes as indicated by the green, red, and purple arrows.
}
\label{fig:SI_Pfaffians}
\end{figure*}

\textit{(3) Explicit calculation of the kernel vector for the RPS chain.} 
With the graph-theoretical definition of the Pfaffian, the adjugate vector $\vec{v}$ in Eq.~\eqref{eq:adjugate_vector_general} for the RPS chain can be directly computed from its network topology; see Fig.~\ref{fig:SI_Pfaffians} for a visualization of the perfect matchings contributing to the components $v_{2m-1}$ and $v_{2m}$. For every component of the adjugate vector, it turns out that only one perfect matching exists on the RPS chain upon removing the relevant component node. Thus, one computes for the adjugate vector by applying Eq.~\eqref{eq:Pfaffian_matchings}:
\begin{align}\label{eq:adjugate_vector}
    \vec{v} = (r_2^{n-1}, r_1 r_2^{n-2}, r_2^{n-2}r_3, r_1 r_2^{n-3} r_3, \dots, r_2^{n-m}r_3^{m-1}, r_1r_2^{n-m-1}r_3^{m-1}, \dots, r_2 r_3^{n-2}, r_1 r_3^{n-2}, r_3^{n-1})^T\ .
\end{align}
Because the adjugate vector is not the zero-vector, the kernel of $A$ is one-dimensional and the adjugate vector spans the kernel. Therefore, upon normalizing with the sum over all entries $C$,
\begin{align}
C = \textstyle\sum_\alpha v_\alpha &= r_2^{n-1}\frac{1-r^{-n}}{1-1/r}+r_3^{n-2}r_1\frac{1-r^{n-1}}{1-r}=r_2^{n-2} \frac{r_2 (1-r^{-n}) + r_1 (1-r^{1-n})}{1-1/r}\ ,
\end{align}
the kernel vector $c_\alpha=v_\alpha/C$ is unique and strictly positive in all components $\alpha$, and can be written as (see Eq.~\eqref{eq:kernel_vector} of the main text): 
\begin{align}\label{eq:kernel vector}
\begin{pmatrix}
c_{2m-1} \\ 
c_{2m}
\end{pmatrix}
= \frac{r_2^{n-m-1}r_3^{m-1}}{C}
\begin{pmatrix}
r_2\\ 
r_1
\end{pmatrix}
= \frac{r_2^{n-2}}{C}
\begin{pmatrix}
r_2\\ 
r_1
\end{pmatrix}
r^{-(m-1)}
= \frac{r_3^{n-1}}{Cr_2}
\begin{pmatrix}
r_2\\ 
r_1
\end{pmatrix}
r^{n-m}\ ,
\end{align}
for $m=1, \dots, n-1$ and $c_{2n-1} = r_3^{n-1}/C= \frac{1}{C}r_2^{n-1}r^{-(n-1)}$, and recall $r=r_2/r_3$. 


%
%

\subsection{Characterization of the stationary state}

Here we supplement the main text with information on the stationary state of the ALVE on the RPS chain. This characterization proceeds along the following three statements: (1) All sites of the RPS chain remain occupied with mass in the stationary state (none of the sites becomes depleted), (2) average masses in the stationary state are given by the strictly positive kernel vector $\vec{c}$ of $A$, and (3) fluctuations of masses in the stationary state are determined by $\vec{c}$ as well.

\textit{(1) All sites of the RPS chain remain occupied with mass in the stationary state.}
We showed in the main text and subsection~\ref{sec:kernel} above that $A$ has an unique strictly positive kernel vector $\vec{c}$ upon normalization ($c_\alpha > 0$ for all $\alpha$ and $A \vec{c} = \vec{0}$, $\sum_\alpha c_\alpha = 1$).
Then, the so-called Kullback-Leibler divergence $D(\vec{c} || \vec{x}) = \sum_\beta c_\beta \ln ( c_\beta / x_\beta)$ of the kernel vector $\vec{c}$ to the masses $\vec x$ is a conserved quantity ($\frac{d}{dt} D(\vec{c}||\vec{x})=\sum_\beta(A\vec{c})_\beta x_\beta=0$) and, thus, $D(\vec{c} || \vec{x})(t) = D(\vec{c} || \vec{x}_0)$ for all times $t$. Due to the conservation of $D$, the mass at every site $\alpha$ remains bounded away from 0 if the dynamics is initialized with $x_\alpha(0)>0$ for all $\alpha$; see also \cite{Knebel2013, Knebel2015, Geiger2018}. In other words, none of the sites becomes depleted of mass.

\textit{(2) Average masses in the stationary state are given by the kernel vector $\vec{c}$ of $A$.}
Let us define the temporal average of the masses, $\langle x_\alpha \rangle_T := 1/T \int_0^T \text{d}t \ x_\alpha$, and consider the quantity $\langle \partial_t \log(x_\alpha)\rangle_T$.
On the one hand,
\begin{align}
    1/T \int_0^T \text{d}t\ \partial_t \log (x_\alpha) = \frac{1}{T}(\log (x_\alpha) (T) - \log (x_\alpha) (0)) \xrightarrow{T \to \infty} 0\ ,
\end{align}
and the convergence to 0 follows because $\log (x_\alpha) (T)$ is bounded, but $1/T$ converges to 0 as $T\to \infty$.
On the other hand, employing the ALVE~\eqref{eq:ALVE} yields,
\begin{align}
    1/T \int_0^T \text{d}t\ \partial_t \log (x_\alpha) = \sum_{\beta} a_{\alpha\beta} \langle x_\beta \rangle_T \ , 
\end{align}
and, thus, $\langle \vec{x} \rangle_T$ converges to the kernel of $A$ as $T\to \infty$. Because,  $\vec{c}$ spans the kernel of $A$, the average masses in the stationary state are given by this kernel vector, that is, $\langle x_\alpha \rangle_\infty = c_\alpha$ for all initial conditions $x_\alpha(0)>0$ for all $\alpha$.\\

\textit{(3) Fluctuations of masses in the stationary state are determined by the kernel vector $\vec{c}$ of $A$.}
In a similar manner as for the average masses, let us consider the quantity $\langle \partial_t x_\alpha\rangle_T$. 
On the one hand,
\begin{align}
    1/T \int_0^T \text{d}t\ \partial_t x_\alpha = \frac{1}{T}(x_\alpha (T) - x_\alpha (0)) \xrightarrow{T \to \infty} 0\ .
\end{align}
On the other hand, upon employing the ALVE~\eqref{eq:ALVE},
\begin{align}
    1/T \int_0^T \text{d}t\ \partial_t x_\alpha = \sum_{\beta} a_{\alpha\beta} \langle x_\alpha x_\beta \rangle_T  \ , 
\end{align}
In other words, $\{\langle x_\alpha x_\beta\rangle_T\}_\beta$ as a function of $\beta$ lies in the kernel of $A$ as $T\to \infty$.
Thus, $\langle x_\alpha x_\beta \rangle_\infty = const (\alpha)\cdot c_\beta$ for $\beta\neq \alpha$ with $\vec{c}$ again denoting the strictly positive kernel vector of $A$. 
With the same arguments one obtains $\langle x_\alpha x_\beta\rangle_T = const(\beta) \cdot c_\alpha$ for $\alpha\neq \beta$ and, thus, $\langle x_\alpha x_\beta\rangle_T \to const \cdot c_\alpha c_\beta$ as $T\to\infty$ for all $\alpha\neq \beta$. 
To compute the second moment of the mass in the stationary state, $\langle x_\alpha^2 \rangle_T\to \langle x_\alpha^2 \rangle_\infty$ as $T\to\infty$, we exploit the normalization $\sum_\alpha x_\alpha=1$ as follows: 
$
    \langle x_\alpha^2 \rangle_T = \langle x_\alpha (1-\sum_{\beta \neq \alpha} x_\beta) \rangle_T 
    = \langle x_\alpha \rangle_T - \sum_{\beta\neq \alpha} \langle x_\alpha x_\beta \rangle_T
$
and, hence, obtain in the stationary state $\langle x_\alpha^2 \rangle_\infty = c_\alpha - const\cdot  c_\alpha \sum_{\alpha \neq \beta} c_\beta$. By exploiting normalization $\sum_{\alpha \neq \beta} c_\beta = (1-c_\alpha)$, the stationary second moment is written as
$
\langle x_\alpha^2 \rangle_\infty = const\cdot c_\alpha^2 +  (1-const) c_\alpha
$. 
Consequently, the stationary state fluctuations ($\text{Var}(x_\alpha)_\infty= \langle x_\alpha^2\rangle_\infty - \langle x_\alpha\rangle_\infty^2$) at site $\alpha$ are quantified as,
\begin{align}\label{eq:variance}
    \text{Var}(x_\alpha)_\infty = \sigma_{\alpha,\infty}^2 =(1-const)c_\alpha(1-c_\alpha)\ .
\end{align}
This functional form was plotted as solid line in the insets of Fig.~\ref{fig:polarization}.
For completeness, the stationary correlations, $\text{Corr}(\alpha,\beta)_\infty= \langle x_\alpha x_\beta\rangle_\infty - \langle x_\alpha\rangle_\infty\langle x_\beta\rangle_\infty$, between sites $\alpha$ and $\beta$ are obtained as,
\begin{align}
    \text{Corr}(\alpha,\beta)_\infty=-(1-const)c_\alpha c_\beta\ .
\end{align}
%


\section{Robustness of polarization on the RPS chain}

\subsection{Robustness against perturbation of the model parameters}

Polarization is robust against perturbations of the model parameters. This characteristics follows from the result that polarization on the RPS chain is a topological phase, but can also be shown directly. Here we supplement the statements from the main text to directly quantify the order of magnitude against which polarization on the RPS chain is robust. To this end, we employ a perturbation of the \textit{positive} matrix entries~\eqref{eq:Matrix_An} as $a_{\alpha\beta}'=a_{\alpha\beta}(1+\epsilon_{\alpha\beta})$; for $a_{\alpha\beta}<0$, the perturbed entry $a_{\alpha\beta}'$ is obtained as $a_{\alpha\beta}' = - a_{\beta\alpha}'$ such that the perturbed matrix $A'$ is antisymmetric. Furthermore, we assume that the perturbations $\epsilon_{\alpha\beta}$ are independently sampled from an identical probability distribution, whose mean fulfils $|\langle\epsilon \rangle| \lesssim 1/n$ and whose probability mass is centered around the mean. All single realizations $\epsilon_{\alpha\beta}$ are assumed to be greater than $-1$ to preserve the network topology of a RPS chain.

For one realization of the perturbed rates, one may compute the adjugate vector of the perturbed system, $\vec{v}^\epsilon$, along the lines leading to Eq.~\eqref{eq:adjugate_vector} and obtains, for example, for the first component $v_1^\epsilon =v_1(1+\epsilon_{32})(1+\epsilon_{54})\dots(1+\epsilon_{S,S-1})$. 
If all realizations of the perturbations $\epsilon_{\alpha \beta}>-1$, then the perturbed component $v_1^\epsilon$ has the same sign as the unperturbed component $v_1$. Consequently, the kernel vector of the perturbed system $A'$ is strictly positive. 

If, in addition, the perturbations are sampled from a probability distribution whose mean fulfills $|\langle\epsilon \rangle| \leq 1/n$, the same polarization pattern emerges as for the unperturbed system if the system size is sufficiently large ($n\gg 1$). This can be seen from the expansion of adjugate vector; for example, the first component scales as $v_1^\epsilon \sim v_1 \big(1 + (n-1)\bar{\epsilon}_{(n-1)}\big)+ \mathcal{O}(v_1n^2 \epsilon_{\alpha_1\beta_1}\epsilon_{\alpha_2\beta_2})$ and $\bar{\epsilon}_{(n-1)} = (\epsilon_{32}+\epsilon_{54}+\dots+\epsilon_{S, S-1})/(n-1)$ denoting the sample average of the perturbations contributing to $v_1$. For large systems, $n\gg 1$, this sample average approaches the mean $\langle \epsilon\rangle$ of the probability distribution. Hence, the corrections to $v_1$ are of the same order of magnitude as $v_1$ because we assumed $|\langle\epsilon \rangle| \lesssim 1/n$. The same arguments hold true for the other components of the adjugate vector of the perturbed system $A'$. Therefore, in total, the same polarization pattern emerges for the perturbed system as for the unperturbed system.

\begin{figure*}[th!]
\centering
\includegraphics[width=1.0\textwidth]{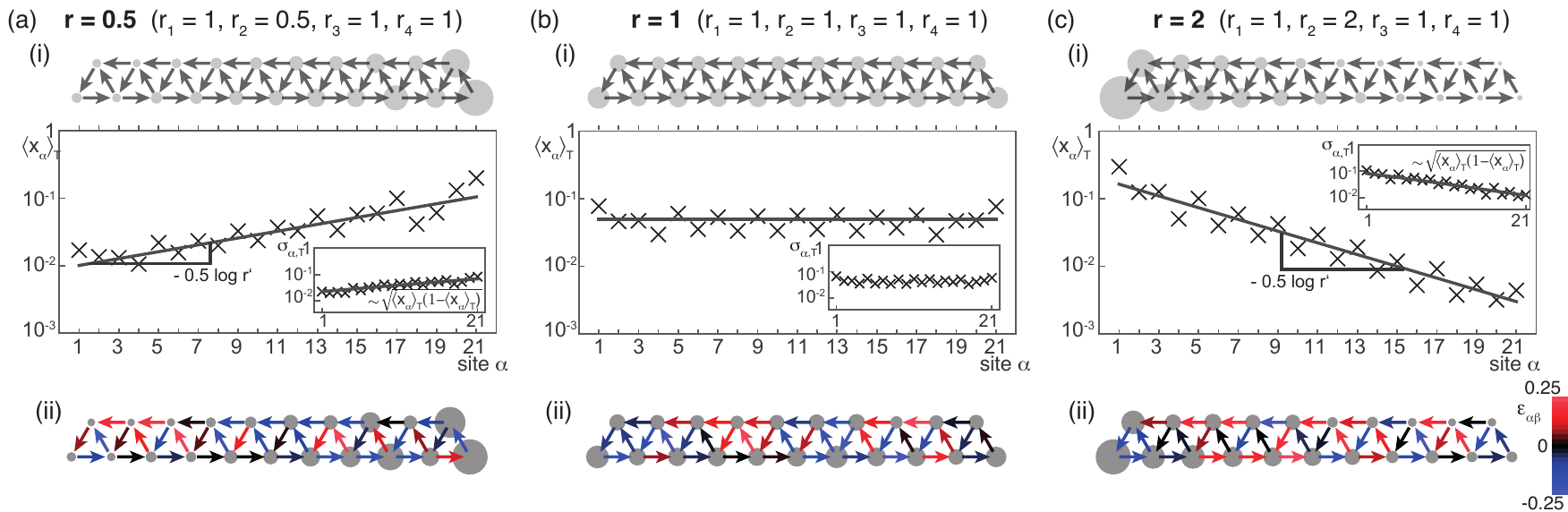}
\caption{\textbf{Polarization of mass to the boundary emerges with the additional coupling $r_4>0$ between even-numbered sites in the same qualitative manner as without this coupling.}
This figure is organized in the same manner as Fig.~\ref{fig:polarization} of the main text.
Mass becomes polarized to the boundary ((a)(i) to the right for $r<1$; (c)(i) to the left for $r>1$) in that average masses, $\langle x_\alpha\rangle_T$, decay exponentially into the bulk at a length scale set by $\ln{r'}$ with a renormalized parameter $r'$ for all initial conditions. For $r=1$, mass becomes  distributed throughout the whole chain (b)(i). Polarization is no state of rest as shown by the non-vanishing fluctuations $\sigma_{\alpha, T}$ around the average masses (insets of graphs (a-c)(i)).
Polarization is robust against perturbations~(a-c)(ii) along the same lines as for $r_4=0$; see Fig.~\ref{fig:polarization}. 
}
\label{fig:SI_polarization_r4}
\end{figure*}

\subsection{Robustness against introducing additional couplings (Fig.~\ref{fig:SI_polarization_r4})}

As mentioned in the main text, we also introduced the additional coupling between even sites. That is, the links between sites $2\alpha$ and $2(\alpha-1)$ for $\alpha= 2, \dots, n-1$ are added to the RPS chain. The results of the simulations are presented in Fig.~\ref{fig:SI_polarization_r4}. Even with these additional couplings, the same qualitative polarization states are approached. 

%
%

\section{Topological band theory of the RPS Hamiltonian}
\subsection{Spectral properties of the RPS Hamiltonian}

\textit{(1) RPS chain of $S+1$ sites with periodic boundary conditions.} 
To make analytical progress within the framework of topological band theory, we extend the RPS chain by one additional site $S+1=2n$ with periodic boundary conditions as described in the main text. 
This way, the corresponding antisymmetric interaction matrix $A_\text{PBC}\in \mathbb{R}^{2n\times 2n}$ extends the antisymmetric matrix $A$ by one row and column. Because of the periodic boundary conditions, $A_\text{PBC}$ is an antisymmetric block-circulant matrix (compare with $A$ in Eq.~\eqref{eq:Matrix_An}):
    \begin{align} \label{eq:Matrix_Cn}
        A_\text{PBC} = \begin{pmatrix}
        A_0 & A_1 & 0 & 0 & \dots & 0 & A_{-1}\\
        A_{-1} & A_0 & A_1 & 0 & \dots & 0 & 0\\
        0 & A_{-1} & A_0 & A_1 & \dots & 0 & 0\\
        \vdots & \vdots & \vdots & \vdots & \ddots & \vdots & \vdots\\
        0 & 0 & 0 & 0 & \dots & A_0 & A_1\\
        A_1 & 0 & 0 & 0 & \dots & A_{-1} & A_0
        \end{pmatrix} = Circ(A_0, A_1, 0, \dots, 0, A_{-1})\ ,
    \end{align}
with the $2 \times 2$ block-matrices:
    \begin{align}\label{eq:block_matrices}
        A_0 = \begin{pmatrix}
        0& r_3 \\ -r_3 & 0
        \end{pmatrix} = -A_0^T, \; 
        A_1 = \begin{pmatrix}
        -r_1 & 0 \\ r_2 & 0
        \end{pmatrix} = -A_{-1}^T\ .
    \end{align}
In this notion, the unit cell of the RPS chain may be interpreted as follows: the transition between sites $2\alpha \to 2\alpha-1$ for $\alpha=1, \dots, n$  (with transition rate $r_3$) define the inter-cellular coupling, encoded by $A_0$; see Fig.~\ref{fig:model}. These unit cells are coupled by the transition rates $r_1$ and $r_2$ with each other and define the intra-cellular couplings, encoded by $A_1=-A_{-1}^T$. 
To connect our analysis to results from condensed matter physics, we defined in the main text the RPS Hamiltonian $H := iA_\text{PBC}$, which is a Hermitian matrix ($H^\dagger = H$) with only real eigenvalues. Accordingly, $H_0 := iA_0$, $H_1 := iA_1$, and $H_{-1} := iA_{-1}$ are defined.\\

\textit{(2) Fourier transform of the RPS Hamiltonian.} 
Because $H$ is a translationally invariant matrix, its spectrum can be computed explicitly. To determine the spectrum of $H$, we start from the eigenvalue equation as follows:
\begin{align}\label{eq:eigenvalue_real_space}
    \lambda^{(\kappa)} u_\alpha^{(\kappa)} = \sum_{\beta=1}^{2n}H_{\alpha\beta} u_\beta^{(\kappa)}\ ,\  \kappa=1, \dots, 2n\ ,
\end{align}
in which $\vec{u}\in \mathbb{R}^{2n}$ is an eigenvector to the eigenvalue $\lambda$, and $\kappa$ labels the different eigenvectors. Please note that the eigenvalues come in pairs of $\pm \lambda$ because $H$ is Hermitian. To exploit translational invariance of $H$ in the spirit of band theory in condensed matter physics, we compute its Fourier transform through the following steps.
First, the ansatz for the eigenvector $\vec{u}$,
\begin{align}\label{eq:eigenvector_real_space}
    u_\alpha^{(\kappa)} = u_{2\alpha' +\alpha''+1}^{(\kappa)} =\tilde{u}^{(\alpha'')}(k)\cdot e^{ik\alpha'}\ ,
\end{align}
is employed with $\alpha = 2\alpha'+\alpha''+1$. This ansatz decomposes the eigenvector into a plane wave part, $e^{ik\alpha'}$, and a within-cell alignment part, $\tilde{u}^{(\alpha'')}(k)$.
Here $\alpha'= 0, \dots, n-1$ labels the unit cell and $\alpha'' = 0,1$ the within-cell position of the lattice site. The momentum $k = \frac{2\pi}{n} l$ is a rescaled version of the index $\kappa$ with $l = -\lfloor \frac{n}{2}\rfloor, \dots, -1, 0, 1, \dots, \lfloor \frac{n}{2}\rfloor-1$.   

Because $H$ is a block-circulant matrix (and, thus, $H$ is translationally invariant), its entries are given as $H_{\alpha \beta} = H_{2\alpha'+\alpha''+1,2\beta'+\beta''+1}=\big(H_{\alpha'-\beta'}\big)_{\alpha'', \beta''}$ with $\alpha'', \beta'' = 0, 1$ and $\alpha', \beta'= 0, \dots, n-1$. Note that only $H_{-1}, H_0, H_1$ are non-zero, see Eq.~\eqref{eq:Matrix_Cn}, and that the blocks $H_l$ are cyclic in the index $l$, that is, $H_l = H_{n+l}$.
With these definitions and properties, the eigenvalue equation~\eqref{eq:eigenvalue_real_space} reduces to:
\begin{align}\label{eq:eigenvalue_momentum_space}
    \lambda(k)\tilde{u}^{(\alpha'')}(k) =\sum_{\beta''=0,1} \tilde{H}(k)_{\alpha'',\beta''}\cdot \tilde{u}^{(\beta'')}(k)\ ,
\end{align}
with $\tilde{H}(k) := \sum_{l=0}^{n-1} H_l e^{i kl}$ as the Fourier transform of the Hamiltonian $H$. 

In summary, the eigenvectors of $H$ as defined in Equation~\eqref{eq:eigenvalue_real_space}  can be decomposed into a plane wave part $e^{ik\alpha'}$ and a polarization part $\tilde{u}^{(\alpha'')}(k)$ (see Equation~\eqref{eq:eigenvector_real_space}). The polarization part fulfills the much simpler eigenvalue equation~\eqref{eq:eigenvalue_momentum_space} for the Fourier transform of the Hamiltonian, which is a $2\times 2$ matrix in this case. The eigenvalues can be directly inferred from Equation~\eqref{eq:eigenvalue_momentum_space} through $\lambda^{(\kappa)} = \lambda(k)$.
The Fourier transform of the Hamiltonian for the RPS chain is obtained as follows (see Eq.~\eqref{eq:Hamiltonian} in the main text):
\begin{align}
    \tilde{H}(k)&=e^{ - i k}  (- H_1^T) + H_0 + e^{i k} H_1\ ,\\
    &= \begin{pmatrix}
2  r_1 \sin k & - r_2 \sin k + ir_3 - ir_2 \cos k \\- r_2 \sin k-ir_3 + ir_2 \cos k   & 0
\end{pmatrix}\ ,\\
&= \begin{pmatrix}
h_0+h_3 & h_1-ih_2 \\
h_1+ih_2   & h_0-h_3
\end{pmatrix}
=
\begin{pmatrix}
r_1 \sin k\\
-r_2 \sin k\\
- r_3 +r_2\cos k \\
r_1 \sin k
\end{pmatrix}\cdot \vec{\sigma}=\begin{pmatrix}h_0(k)\\h_1(k)\\h_2(k)\\h_3(k)\end{pmatrix}\cdot \vec{\sigma}\ 
= \vec{h} \cdot \vec{\sigma}\ ,
\end{align}
where, in the last line, we used the compact notation $\vec{\sigma} = (\sigma_0, \sigma_1, \sigma_2, \sigma_3)^T$ of the Pauli matrices:
\begin{align}
\sigma_0 = \begin{pmatrix}
1 & 0 \\ 0 &1
\end{pmatrix} \; \sigma_1 = \begin{pmatrix}
0 & 1 \\ 1 & 0
\end{pmatrix} &\; \sigma_2 = \begin{pmatrix}
0 & -i \\ i & 0
\end{pmatrix}\; \sigma_3 = \begin{pmatrix}
1 & 0 \\ 0 & -1
\end{pmatrix}\ .\\\nonumber
\end{align}

\textit{(3) Symmetries of the RPS Hamiltonian.}
$\tilde{H}$ has an inherent ``particle hole symmetry'' as mentioned in the main text. With the unitary operator $\mathcal{C}=\sigma_0\circ \kappa$, where $\sigma_0$ is the identity matrix and $\kappa$ is the complex conjugation, this symmetry is understood in the sense that $\sigma_0\circ \kappa \tilde H(k) + \tilde H(-k)\sigma_0\circ \kappa = \kappa \tilde H(k) + \tilde H(-k) = 0$ (as an operator equality) if applied to some matrix. This particle-hole symmetry of $\tilde{H}$ originates from the definition of the RPS Hamiltonian $H=iA_\text{PBC}$ via the real-valued antisymmetric matrix $A$ and the definition of the Fourier transform. As a consequence, the spectrum of $\tilde{H}$, and thus $H$, is point-symmetric to the origin, $\lambda(k) = -\lambda(-k)$; see Fig.~\ref{fig:spectrum}.\\

\textit{(4) Spectrum of the RPS Hamiltonian.}
For completeness, we briefly mention explicitly the spectral properties of the RPS Hamiltonian and its Fourier transform. 
The determinant of $\tilde{H}$ is given by:
\begin{align}
\label{eq:H_determinant}
\begin{split}
\det \tilde H(k) &= h_0(k)^2 - h_1(k)^2 - h_2(k) ^2 - h_3(k) ^2=  -\big(r_2^2 \sin^2 k + (r_3 - r_2 \cos k)^2 \big)\leq 0\ .
\end{split}
\end{align}
Therefore, the determinant is 0 only if $r_2 = r_3$ (that is, $r=1$) and at $k = 0$. In other words, $\tilde{H}$ has a non-trivial kernel only if $r=1$. The trace of $\tilde{H}$ is given by $\text{tr} (\tilde{H}) = 2 h_0$.
Thus, the eigenvalues of $\tilde{H}$ are obtained in two bands $\lambda^{+}$ and $\lambda^{-}$ as follows:
\begin{align}
\lambda^{\pm}(k) 
&= \text{tr} (\tilde{H})/2 \pm \sqrt{(\text{tr} (\tilde{H}) )^2/4 - \det \tilde{H}}
= h_0(k) \pm \sqrt{h_1(k)^2+h_2(k)^2+h_3(k)^2} \ , \\
&= r_1 \sin k \pm\sqrt{(r_1^2 + r_2^2)\sin^2 k +(r_3-r_2\cos k)^2}\ .
\end{align}
The value of $r_1$ only affects the eigenvalues of $\tilde{H}$ as a global shift and this shift vanishes for $k = 0$. As already seen from the determinant, the eigenvalues are only 0 for $r=1$ at $k=0$. \\

\textit{(5) Eigenvectors of the RPS Hamiltonian.}
The eigenvectors, $\tilde{\vec{u}}^\pm(k)$, of $\tilde{H}$ corresponding to $\lambda^\pm(k)$, $k$ real-valued, can be written as follows (not normalized):
\begin{align}
\tilde{\vec{u}}^{\pm }(k)&=  
\begin{pmatrix}
\lambda^{\pm}(k)  - h_0(k) + h_3(k)\\ h_1(k) + i h_2 (k)
\end{pmatrix} 
=
 \begin{pmatrix}
h_3(k)\pm \sqrt{h_1(k)^2+h_2(k)^2+h_3(k)^2}\\ h_1(k) + i h_2(k) 
\end{pmatrix}\ ,\\
&= 
\begin{pmatrix}
    r_1\sin k \pm \sqrt{ (r_1^2 + r_2^2) \sin^2k + (r_3 -  r_2 \cos k)^2 }\\
    - r_2 \sin k + i (r_3 - r_2 \cos k)
    \end{pmatrix}\ .\\\nonumber
\end{align}

\textit{(6) Heuristic bulk-boundary correspondence.}
The RPS Hamiltonian $\tilde{H}$ has eigenvalues $\lambda^{\pm}(k)$ for only real $k$ values; see above. However, for the \textit{imaginary} values $k_\pm=\pm i \ln r$, that is $e^{ik_{+}} = 1/r$ and $e^{ik_{-}}=r$, one checks that $\det \tilde{H}(k_\pm) = 0$ (equivalently, $\lambda^{-}(k_\pm) = 0$) as if $k_\pm$ would give rise to additional zero eigenvalues of $\tilde{H}$ (and, thus, $H$). Indeed, it turns out that the value of $k_+=i\ln r$ gives rise to the kernel vector of the RPS chain defined by the antisymmetric matrix $A$ in Eq.~\eqref{eq:Matrix_An}. The correspondence between the ``suppressed'' zero eigenvalue of $H$ defined by the imaginary momentum $k_+$ and the realized zero eigenvalue and corresponding eigenvector of $A$ is governed by the so-called bulk-boundary correspondence. Below, we supplement the main text with a heuristic picture of how the kernel vector of $A$ emerges from the ``suppressed'' eigenvector corresponding to $\lambda^{-}(k_+)$.
The mathematical details of this connection between the periodic RPS chain, defined by $A_\text{PBC}$ and $H$, and the finite RPS chain, defined by $A$, are outlined in the main text and described further below.

The corresponding vectors $\tilde{\vec{w}}^{\pm}$ fulfilling the eigenvalue equation of $\tilde{H}$ for the values of $k_\pm$ are obtained as a solution to the equation $\tilde{H}(k_\pm)\tilde{\vec{w}}^{\pm}= \vec{0}$ as follows:
\begin{align}
\tilde{\vec{w}}^{+}(k_{+} = i \ln r)=  
\begin{pmatrix}
r_2\\ r_1
\end{pmatrix}\ \ \text {, and}\ \
\tilde{\vec{w}}^{-}(k_{-} = -i \ln r)=  
\begin{pmatrix}
0\\ 1
\end{pmatrix}\ .
\end{align}
Only the vector $\tilde{\vec{w}}^+$ is strictly positive and may give rise to a strictly positive kernel vector of $A$; see further below. If $\tilde{\vec{w}}^+$ is treated as an eigenvector of $\tilde{H}$, one can anticipate the polarization behavior of the RPS chain from the form of the corresponding eigenvector $\vec{w}^+$ of $H$ that was introduced in Eq.~\eqref{eq:eigenvector_real_space}:
\begin{align}\label{eq:suppressed_kernel_vectors}
\begin{pmatrix}
w^+_{2\alpha'}\\ w^+_{2\alpha'+1}
\end{pmatrix}
=
\begin{pmatrix}
r_2\\ r_1
\end{pmatrix}
\cdot e^{ik_{+} \alpha'}
=
\begin{pmatrix}
r_2\\ r_1
\end{pmatrix}
\cdot r^{-\alpha'} \ ,\  \text{for } \alpha' = 0, \dots, n-1\ ,
\end{align}
in which the same functional form as the algebraically calculated kernel vector in Eq.~\eqref{eq:kernel_vector} of the main text is apparent.
In other words, the plane wave part $e^{ik\alpha'}$ with $k_+$ becomes the polarization part $e^{-\ln r \cdot \alpha'} \sim e^{-\ln r  \cdot \alpha/2} = e^{-\alpha/l_p}$ with the penetration depth $p_l=2/\ln r$ as observed in the simulations, see main text and  Fig.~\ref{fig:polarization}. It also follows that for $r>1$ mass will get polarized to the left, whereas for $r<1$ polarization occurs to the right boundary of the RPS chain. \\


\textit{(7) Projection of the vector $\vec{h}(k)$ in Fig.~\ref{fig:spectrum}.}
In Fig.~\ref{fig:spectrum}, we plotted the projection of the vector $\vec{h}(k)$, which characterizes the Fourier-transformed Hamiltonian $\tilde{H}$, to analyze the manifold that is mapped out by its eigenvectors as a function of $k$. Because $h_0$ does not contribute to the form of the eigenvectors, only the three components $h_1(k),h_2(k),h_3(k)$ need to be considered for this discussion. The curve traced out by the vector $(h_1(k), h_2(k), h_3(k))=(-r_2\sin k, -r_3 + r_2 \cos k, r_1 \sin k)$ for $k \in [-\pi, \pi)$ lies in the plane with normal vector $\vec{n}=(r_1, 0, r_2)$ as one checks that $\vec{n}\cdot (h_1(k), h_2(k), h_3(k))=0$ for all~$k$.
Thus, instead of plotting $(h_1(k), h_2(k), h_3(k))$, we plotted the first two components of the  vector $(h^{\parallel}(k), h_2(k), 0)$ in that plane defined by $\vec{n}$. This projection is obtained via the rotation matrix $T\in \mathbb{R}^{3\times 3}$ ($T^T T =T T^T= \mathbb 1$):
\begin{align}
    T= \begin{pmatrix}
- \frac{r_1}{\sqrt{r_1^2 + r_2^2}} & 0 & \frac{r_2}{\sqrt{r_1^2 + r_2^2}} \\
0 & 1 & 0\\
 \frac{r_1}{\sqrt{r_1^2 + r_2^2}} & 0 & \frac{r_2}{\sqrt{r_1^2 + r_2^2}} 
\end{pmatrix}\ ,
\end{align}
applied to the vector $(h_1(k), h_2(k), h_3(k))$:
\begin{align}
\begin{pmatrix}
h^{\parallel}(k) \\ h_2(k) \\ 0
\end{pmatrix} 
&:=T\cdot
\begin{pmatrix}
h_1(k) \\ h_2(k) \\ h_3(k)
\end{pmatrix} 
=  
\begin{pmatrix}
- \frac{r_1}{\sqrt{r_1^2 + r_2^2}} & 0 & \frac{r_2}{\sqrt{r_1^2 + r_2^2}} \\
0 & 1 & 0\\
 \frac{r_1}{\sqrt{r_1^2 + r_2^2}} & 0 & \frac{r_2}{\sqrt{r_1^2 + r_2^2}} 
\end{pmatrix}
\cdot
\begin{pmatrix}
-r_2 \sin k\\ -r_3 + r_2 \cos k\\ r_1 \sin k
\end{pmatrix}  = 
\begin{pmatrix}
 \frac{2 r_1 r_2 }{\sqrt{r_1^2 + r_2^2}} \sin k
\\ -r_3 + r_2 \cos k\\ 0
\end{pmatrix}\ .
\end{align}


\subsection{Explicit bulk-boundary correspondence}

\begin{figure*}[ht!]
\centering
\includegraphics[width=1.\textwidth]{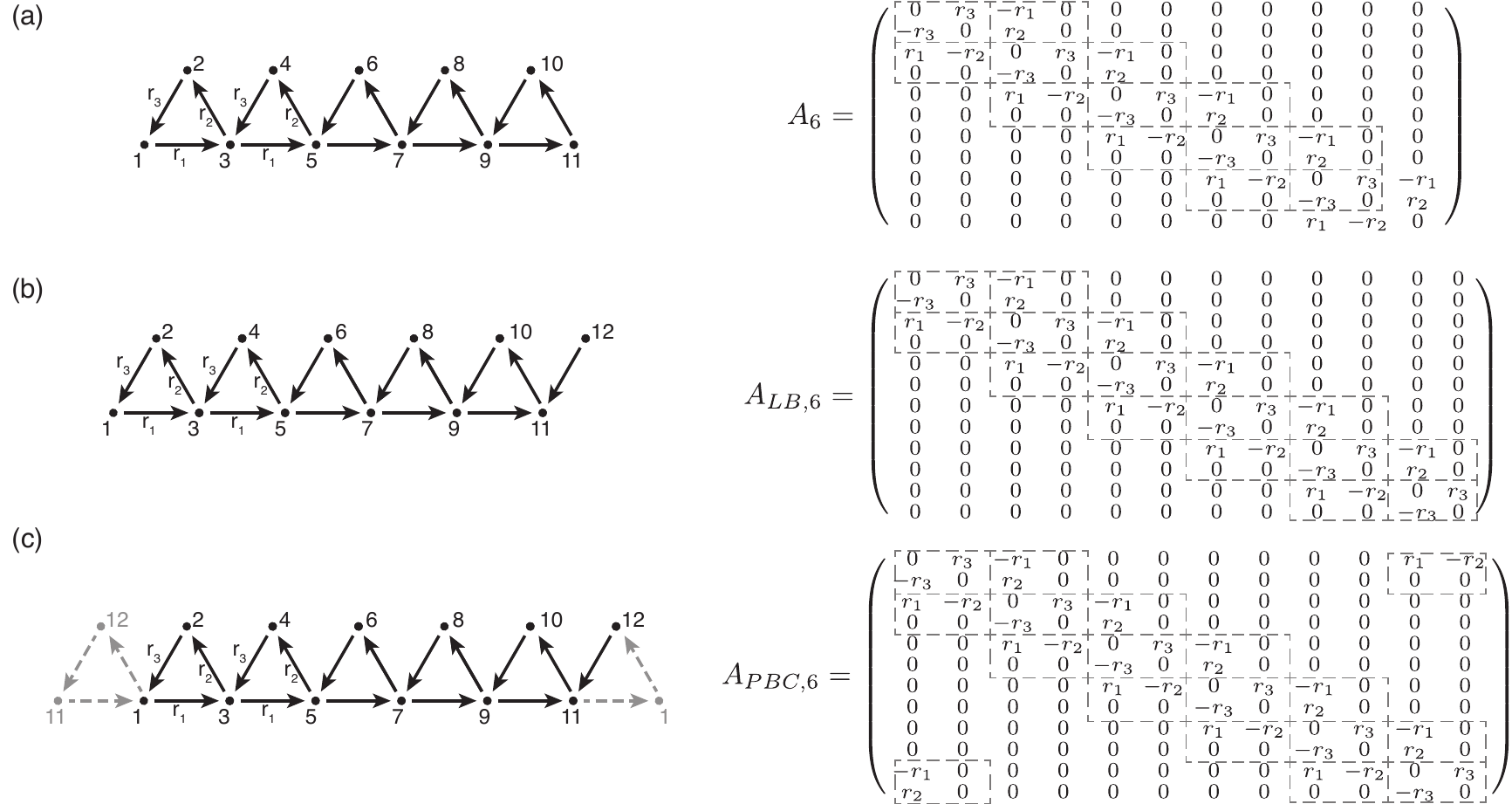}
\caption{\textbf{Visualization of the three different set-ups of the RPS chain, exemplified for $n=6$.}
(a) RPS chain on $S=2n-1$ sites defined by the antisymmetric matrix $A_n$. 
(b) Left-boundary RPS chain on $S+1=2n$ sites defined by the antisymmetric matrix $A_{\text{LB},n}$, which is a block-Toeplitz matrix~\cite{Gray2006}. 
(c) Periodic RPS chain on $S+1=2n$ sites with periodic boundary conditions defined by the antisymmetric matrix $A_{\text{PBC},n}$, which is a block-circulant matrix.
}
\label{fig:SI_boundaries}
\end{figure*}

Above, we heuristically explained that a ``suppressed'' zero eigenvalue of the RPS Hamiltonian $H$, or equivalently $A_\text{PBC}$, becomes a zero eigenvalue of $A$ with a corresponding strictly positive kernel vector. Here we make this heuristic explanation rigorous by introducing an intermediary matrix between the block-circulant matrix $A_\text{PBC}$ and $A$, and apply the so-called Szegö-Widom theorem~\cite{Deift2012,Basor2019}.
The three different systems that we discuss here are visualized in Fig.~\ref{fig:SI_boundaries} and are summarized as follows (we use the subscript $n$ in the matrices from now on to highlight the system size):
\begin{itemize}
    \item \textit{The RPS chain} on $S=2n-1$ sites, see Fig.~\ref{fig:SI_boundaries}(a); defined by the antisymmetric matrix $A_n$ in Eq.~\eqref{eq:Matrix_An}.
    \item \textit{The left-boundary RPS chain} on $S+1=2n$ sites, but both sites $S$ and $S+1$ are not connected to site 1, that is, this system has the same left boundary as the RPS chain (thus, ``LB'' as subscript), but a different right boundary, see Fig.~\ref{fig:SI_boundaries}(b); defined by the antisymmetric matrix $A_{\text{LB},n}$, which is a block-Toeplitz matrix~\cite{Gray2006}. This block-Toeplitz matrix represents the bridge to analyze the boundary properties of the RPS chain starting from the properties of the periodic RPS chain. 
%
    \item \textit{The periodic RPS chain} on $S+1=2n$ sites with periodic boundary conditions (sites $S$ and $S+1$ are also connected to site 1 in a rock-paper-scissors cycle), see Fig.~\ref{fig:SI_boundaries}(c); defined by the antisymmetric matrix $A_{\text{PBC},n}$ in Eq.~\eqref{eq:Matrix_Cn}, which is a block-circulant matrix, and represents the bulk properties of the RPS chain. 
\end{itemize}

The goal is to explain the emergence of a strictly positive kernel vector of $A_n$ with polarization to the boundary, which follows from the bulk properties of the RPS chain $A_{\text{PBC},n}$. To this end, we proceed in four steps:
\begin{enumerate}
    \item [(1)]
 We introduce the left-boundary RPS chain, defined by $A_{\text{LB}, n}$. We show that, by virtue of the Szegö-Widom theorem, two eigenvalues approach 0 as $n\to \infty$ if $r>1$. In other words, the left-boundary RPS chain has two ``asymptotic'' zero eigenvalues for $r>1$, which represents the essence of the bulk-boundary correspondence. 
  \item [(2)]
We show that the vector $\vec{u}^+$ as obtained in Eq.~\eqref{eq:suppressed_kernel_vectors}, which fulfils the eigenvalue equation of $H$ to the ``suppressed'' zero eigenvalue at value $k=i\ln r$, is indeed such an asymptotic kernel vector of $A_{\text{LB}, n}$.
 \item [(3)]
This strictly positive vector $\vec{u}^+$ gives rise to the strictly positive kernel vector of $A_n$, which is unique upon normalization and reflects polarization to the left. 
 \item [(4)]
The same arguments can be carried out for a right-boundary RPS chain, defined by $A_{\text{RB}, n}$, showing that the same  vector $\vec{u}^+$ gives rise to the strictly positive kernel vector of $A_n$. This time, however, polarization to the right for $r<1$ follows.
\end{enumerate}

\begin{figure*}[ht!]
\centering
\includegraphics[width=1.0\textwidth]{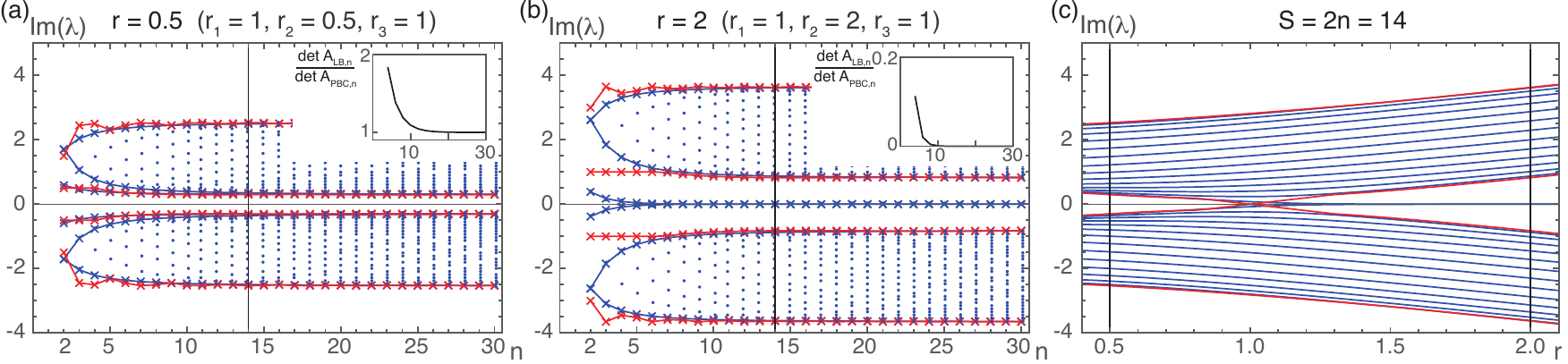}
\caption{\textbf{Bulk-boundary correspondence becomes apparent for the left-boundary RPS chain in that two asymptotic zero eigenvalues emerge as the system size $n\to\infty$ if $r>1$.}
(a) Imaginary part of the eigenvalues of $A_{\text{LB},n}$ (for $r=0.5<1$) indicated by blue dots depending on the system size $S=2n$. Largest and smallest eigenvalues above and below 0 are depicted as solid lines; red line indicates largest and smallest eigenvalues of $A_{\text{PBC},n}$.
The spectral gap of the Toeplitz matrix remains the same as for periodic RPS chain such that $\det(A_{\text{LB},n})/\det (A_{\text{PBC},n})\to 1$ as $n\to \infty$; see inset.
(b) Same plot as for (a) but with parameter $r=2>1$. Two eigenvalues appear in the spectral gap and approach 0 as the system size increases such that $\det(A_{\text{LB},n})/\det (A_{\text{PBC},n})\to 0$ as $n\to \infty$. Thus, the behavior of the $\det(A_{\text{LB},n})$ is qualitatively different from the circulant matrix $A_{\text{PBC},n}$ in that two asymptotic kernel vectors arise for the left-boundary RPS chain. One of these kernel vectors gives rise to the polarization state of the RPS chain.
(c) Spectrum of the Toeplitz matrix for system size $n=14$ (indicated as black vertical lines in (a) and (b)) for different values of the control parameter $r$ ($r_1=r_3=1$, $r_2$ is varied; $r=0.5$ and $r=2$ are depicted in (a) and (b), indicated by black vertical lines in (c)). For $r<1$, all eigenvalues of the Toeplitz matrix (blue lines) lie between the largest and smallest eigenvalues of the circulant matrix (red lines) above and below 0. For $r>1$, the circulant matrix again has a spectral gap, whereas the Toeplitz matrix has a pair of eigenvalues that approach zero as $n\to \infty$.}
\label{fig:SI_szegowidom}
\end{figure*}

\textit{(1) Asymptotic zero eigenvalues for the left-boundary Toeplitz matrix for $r>1$.} In the spectrum of the Toeplitz matrix $A_\text{LB}$, a pair of zero eigenvalues emerges as $n$ increases for $r>1$ as opposed to the gapped spectrum of the circulant matrix $A_\text{PBC}$, whereas for $r<1$ the spectra between $A_\text{LB}$ and $A_\text{PBC}$ develop similarly; see Fig.~\ref{fig:SI_szegowidom}.

More precisely, one may apply the so-called Szegö-Widom theorem for block-Toeplitz matrices, as detailed in reference~\cite{Basor2019}, to the matrix $A_{\text{LB},n}$. From this theorem it follows that the ratio between the determinants of the left-boundary RPS chain and the periodic RPS chain, $\det(A_{\text{LB},n})/\det (A_{\text{PBC},n})$, approaches a constant $E(\tilde{H})$, which only depends on the Fourier transform $\tilde H$, as $n\to \infty$. 
If $E$ is non-zero, both the spectra of $A_{\text{PBC},n}$ and $A_{\text{LB},n}$ behave similarly and $A_{\text{LB},n}$ is well approximated by $A_{\text{PBC},n}$, loosely speaking. However, when $E = 0$, the spectra of $A_{\text{PBC},n}$ and $A_{\text{LB},n}$  differ qualitatively, and this case is linked with topologically non-trivial behavior~\cite{Basor2019,Dubail2017}. 

For the RPS chain, the ratio of the two determinants can be explicitly computed. For example by applying the graph-theoretical formulation of the Pfaffian, the Pfaffian of both the block-Toeplitz and the block-circulant matrix, $A_{\text{LB},n}$ and $A_{\text{PBC},n}$, are calculated as follows:
\begin{align}
\text{Pf} (A_{\text{LB},n}) &= 
\text{sign} \big(
\begin{pmatrix}
1 &2& 3& 4& ...& S& S+1\big)
\end{pmatrix} 
r_3^{n} 
= (+1)r_3^n\ ,\\
\text{Pf} (A_{\text{PBC},n}) 
&=  
\text{sign} \big(
\begin{pmatrix}
    1 &2 &3 &4 &...& S& S+1
\end{pmatrix} \big)
r_3^{n} + 
\text{sign}\big( 
\begin{pmatrix}
    S+1 & 1 &2 &3 &4 &...& S
\end{pmatrix}\big)
r_2^{n} 
= r_3^n - r_2^n \ .
\end{align}

Thus, one obtains for the fraction of the determinants:
\begin{align}
    \frac{\det A_{\text{LB},n}}{\det A_{\text{PBC},n}} &= \frac{r_3^{2n}}{(r_3^n - r_2^n )^2} = \frac{1 }{(1- r^n)^2} \xrightarrow{n\to \infty} \begin{cases}
    0 & \text{for } r>1 \\
    1 & \text{for } r < 1\ .
    \end{cases}
\end{align}
In the topologically trivial phase ($\mathcal M = +1$, $r<1$), one finds that the determinants behave asymptotically qualitatively similar, whereas in the topologically non-trivial case ($\mathcal M = -1$, $r>1$), one encounters the case $E=0$ and the determinants differ in their qualitative behavior as $n\to \infty$.
Notably, while the determinant of the Toeplitz matrix, $\det A_{\text{LB},n} = r_3^{2n}$, is always nonzero, with increasing size $n$, a pair of eigenvalues approaches zero. Thus, the corresponding eigenvectors become asymptotic kernel vectors of $\det A_{\text{LB},n}$ as $n\to \infty$; see Fig.~\ref{fig:SI_szegowidom}. \\

\textit{(2) Asymptotic kernel vector of the left-boundary Toeplitz matrix for $r>1$.} To determine the asymptotic kernel vectors, one may check the vector $\vec{u}^+$ of the RPS Hamiltonian in Eq.~\eqref{eq:suppressed_kernel_vectors}. This vector is a promising candidate for an asymptotic kernel vector because it fulfils the eigenvalue equation for $H$ at an imaginary value of $k_+= i \ln r$ for which $\lambda(k_+) = 0$ and is a strictly positive vector. 
To this end, one computes:
\begin{align}
A_{\text{LB},n}\vec{u}^+ = 
\begin{pmatrix}
0 & r_3 & -r_1 & 0 &  \dots& 0 & 0\\
-r_3 & 0 & r_2 & 0 & \dots& 0 & 0\\
r_1 & -r_2 & 0 & r_3 & \dots& 0 & 0\\
0 & 0 & -r_3 & 0 & \dots& 0 & 0\\
\vdots & \vdots& \vdots &\vdots & \ddots&\vdots&\vdots\\
0&0&0&0&\dots&0&0\\
0&0&0&0&\dots&0&r_3\\
0&0&0&0&\dots&-r_3&0
\end{pmatrix} 
\cdot
\begin{pmatrix}
r_3 \\ r_1r_3/r_2 \\ r_3^2/ r_2 \\ r_1r_3^2/r_2^2\\
\vdots\\
r_1r_3^{n-1}/r_2^{n-1}\\
r_3^n/r_2^{n-1}\\
r_1r_3^n/r_2^n
\end{pmatrix}
&= \begin{pmatrix}
0\\
0\\
0\\
0\\
\vdots\\
0\\
r_3r_1/r^n\\
-r_3^2/r^{n-1}
\end{pmatrix} \ .
\end{align}
Indeed, if $r>1$, the ``suppressed'' kernel vector $\vec{u}^+$ of $H$ is an ``asymptotic'' kernel vector of the Toeplitz matrix, that is $A_{\text{LB},n}\vec{u}^+ \to \vec{0} $ as $n\to \infty$. In particular, $\vec{u}^+$ is strictly positive.
Note that the second asymptotic kernel vector of $A_{\text{LB},n}$ is not provided by the vector $\vec{u}^-$ because $A_{\text{LB},n}\vec{u}^- = (r_3, 0, 0, \dots, 0)$ does not approach $ \vec{0}$ as $n\to \infty$.
\\

\textit{(3) The asymptotic kernel vector gives rise to the left-polarization kernel vector of the RPS chain for $r>1$.} 
Importantly, $\vec{u}^+$ gives rise to the single kernel vector of $A$, if site $S+1$ is removed from the left-boundary RPS chain to obtain back the original RPS chain of $S$ sites. This way, one obtains the matrix $A_n$ from $A_{\text{LB},n}$ by removing the last column and row. Similarly, the vector that is obtained from $\vec{u}^+$ by removing the last entry is the unique kernel vector of $A_n$ upon normalization. This kernel vector reflects polarization of mass to the left boundary for the case $r>1$.\\

\begin{figure*}[ht!]
\centering
\includegraphics[width=1.0\textwidth]{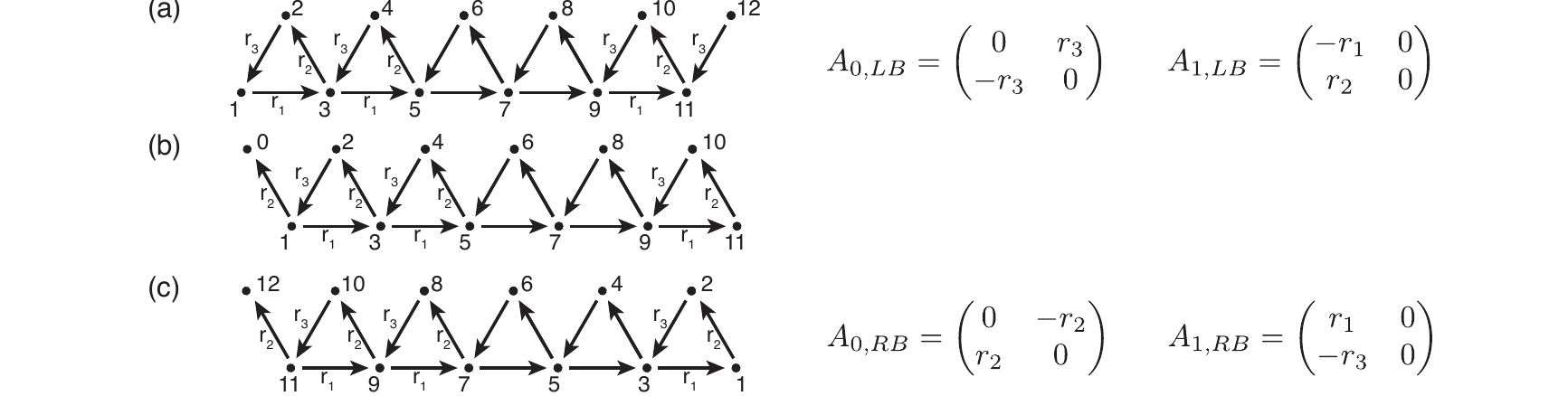}
\caption{\textbf{Toeplitz matrices of the left-boundary and the right-boundary RPS chain.}
(a) Left-boundary RPS chain, defined by $A_{\text{LB},n}$ through the block matrices $A_{0,\text{LB}}$ and $A_{1,\text{LB}}$; see Eq.~\eqref{eq:block_matrices}. 
(b) Right-boundary RPS chain.
(c) Right-boundary RPS chain with relabeled lattices sites such that the same analysis as for the left-boundary RPS chain can be applied. This system is defined by defined by $A_{\text{RB},n}$ through the block matrices $A_{0,\text{RB}}$ and $A_{1,\text{RB}}$ such that the roles between $r_2$ and $r_3$ are swapped and an overall minus sign is obtained.}
\label{fig:boundary_chains}
\end{figure*}

\textit{(4) The vector $\vec{u}^+$ gives rise to the right-polarization kernel vector of the RPS chain for $r<1$ if the right-boundary RPS chain is considered.}
The same arguments (1)-(3) can be made for a right-boundary system as depicted in Fig.~\ref{fig:boundary_chains}(b). Instead of implementing the correct boundary on the left side of the RPS chain (left-boundary RPS chain, defined $A_{\text{LB},n}$; see Fig.~\ref{fig:boundary_chains}(a)), one may equally well implement the correct boundary at the right side of the RPS chain (see Fig.~\ref{fig:boundary_chains}(b)). After relabeling of the lattice sites $\alpha \mapsto 2n-\alpha$, one obtains the right-boundary RPS chain, defined by $A_{\text{RB},n}$; see Fig.~\ref{fig:boundary_chains}(c). This relabeling has the advantage that the analysis above for the left-boundary RPS chain can be applied directly to the right-boundary RPS chain. The difference between $A_{\text{RB},n}$ and $A_{\text{LB},n}$ is that the roles between the rates $r_3$ and $r_2$ are swapped and that $A_{\text{RB},n}$ carries an overall minus sign as compared with $A_{\text{LB},n}$. 

With these preparatory steps one can apply all steps of the above analysis for the left-boundary system to the right-boundary system. It follows that the same  vector $\vec{u}^+$ gives rise to the strictly positive kernel vector of $A_n$, this time, however for the case $r_3/r_2 = 1/r >1$, that is $r<1$. Since the labeling starts with site 1 from the right, one also obtains polarization to the right instead of polarization to the left. 
In summary, the bulk-boundary correspondence yields polarization to the left boundary of the RPS chain if $r>1$ and to the right if $r<1$.

\end{document}